\documentclass[%
 reprint,
 amsmath,amssymb,
 aps,
pra,
]{revtex4-1}

\usepackage{graphicx}
\usepackage{dcolumn}
\usepackage{bm}

\usepackage[normalem]{ulem}

\def\sec#1{Sec.\ \ref{#1}}
\def\eq#1{Eq.\ (\ref{#1})}
\def\fig#1{Fig.\ \ref{#1}}

\begin{document}


\title{Oscillations in electron transport caused by multiple resonances \\ in
       a quantum dot-QED system in the steady-state regime}
\thanks{A footnote to the article title}%

\author{Nzar Rauf Abdullah}
\email{nzar.r.abdullah@gmail.com}
\altaffiliation[]{Physics Department, College of Science, 
             University of Sulaimani, Kurdistan Region, Iraq}
\affiliation{
             Physics Department, College of Science, 
             University of Sulaimani, Kurdistan Region, Iraq
             }%
\affiliation{
            Komar Research Center, Komar University of 
            Science and Technology, Sulaimani City, Iraq
            }%
\author{Chi-Shung Tang}
\affiliation{
             Department of Mechanical Engineering, National United University, 
             1, Lienda, Miaoli 36003, Taiwan. 
}%
\author{Andrei Manolescu}
\affiliation{%
             Reykjavik University, School of Science and Engineering,
             Menntavegur 1, IS-101 Reykjavik, Iceland. 
             }%
\author{Vidar Gudmundsson}
\affiliation{%
             Science Institute, University of Iceland,
             Dunhaga 3, IS-107 Reykjavik, Iceland.
             }%



\begin{abstract}
We model the electron transport current as the photon energy is swept through several resonances of 
a multi-level quantum dot, embedded in a short quantum wire, coupled to photon cavity. 
We use a Markovian quantum master equation appropriate for the long-time evolution and 
include the electron-electron and both the para- and the diamagnetic electron-photon interactions  
via diagonalization in a truncated many-body Fock space.
Tuning the photon energy, several anti-crossings caused by Rabi-splitting
in the energy spectrum of the quantum dot system are found. 
The strength of the Rabi-splittings and the photon-exchange between the resonant states depend 
on the polarization of the cavity photon field. 
We observe oscillations of the charge in the system and several resonant transport current peaks 
for the photon energies corresponding to the resonances in the steady-state regime. 
\begin{description}
%
%
\item[PACS numbers]
42.50.Pq, 05.60.Gg, 73.21.La, 71.38.−k
%
%
\end{description}
\end{abstract}

\pacs{Valid PACS appear here}
\maketitle


\section{Introduction}\label{Sec:Introduction} 

Quantum dot (QD) is a small island that can be connected to electron reservoirs via a tunneling 
region for the sake of studying electron transport~\cite{Kouwenhoven_2001}.
The properties of electron transport 
through the energy levels of a QD are determined by the bias voltage of the electron reservoirs
and the potential profile of the coupling region or tunneling region.
The energy required for making transport from the electron reservoirs to the QD is determined 
by a single electron quantization and the Coulomb interaction~\cite{PhysRevB.53.12625}. 
Therefore, the Coulomb interaction plays an important role in the electron transport through a QD in which 
the maximum resistance can be observed in the presence of an electron-electron interaction bringing forward 
the Coulomb blockade regime~\cite{PhysRevLett.20.1504,Wharam1995}.

The electron transport through the Coulomb blockade QD can be enhanced when 
the QD is coupled to a microwave radiation~\cite{PhysRevLett.73.3443}. 
The transport properties of a QD coupled to a microwave radiation are interesting because of the coexistence role
of the electron and the photon~\cite{Oosterkamp1998}.
In addition, the QD provides a perfect mechanism for the interplay of discrete 
quantum states with the photon~\cite{PhysRevB.78.125308}. 
Therefore, the QD embedded in a photon cavity has been found to be one of the fascinating physical system 
for quantum information process~\cite{PhysRevLett.83.4204,PhysRevLett.85.5647} and 
quantum connectivity in networks~\cite{Kimble2008}.

In addition, several quantum confined geometries to characterize the
influence of photons on the features of electron transport have been studied. 
For example, a quantum wire for exploring the electron population inversion~\cite{Niu56.R12752}, 
a quantum ring-dot for investigating quantum charge pumping~\cite{Torres72.245339}, 
and a quantum point contact including photon-induced intersubband transitions~\cite{Hu62.837}.
In double quantum dots coupled to electromagnetic irradiation, spin-filtering effect~\cite{Watzel99.192101} and 
two types of photon-controlled electron transport related to the ground state and 
excited state resonances~\cite{Shibata109.077401} are shown.

Another aspect of quantum response for investigating transition of electrons between energy levels of 
a quantum dot under the photon field is the thermal broadening of the Fermi level 
in the electron reservoir and the broadening of the confined QD levels~\cite{Ishibashi314.437}.
In order to study the quantum features of a resonance energy level under the photon field, the energy spacing of a quantum system 
should be greater than the thermal energy $\Delta E \gg \kappa_B T$. Therefore, 
the impact of intraband excitation can be seen and the intraband transition can efficiently 
influence electron transport~\cite{Uwe_2004, doi:10.1080/15567260802591985,PhysRevB.71.235302}. 
This process has been observed experimentally~\cite{Wiel_2006} and theoretically~\cite{doi:10.1063/1.4867099}
in QD coupled to microwave radiation which in turns can be used in fluorescence applications~\cite{Monton2009}.
The intraband transitions and the effects of resonance energy levels of a two level system coupled to 
quantized photon field have been used to describe several aspects of the interaction between light and matter 
in solid-state electronic nanodevices~\cite{PhysRevLett.92.127404} with quantum-optical spectroscopy~\cite{Kreinberg2018}.

In our previous works, we studied time-dependent 
photon-assisted electron transport through a single~\cite{Nzar.25.465302} 
and double-quantum dots~\cite{PhysicaE.64.254, Vidar:ANDP201500298} where both electronic systems are coupled 
to photons in the cavity. It was shown that the photon replica states are formed in the presence of the photon field 
and these replicas states in both resonant and off-resonant regimes can have a big role in electronic~\cite{ABDULLAH2016280} and 
thermal~\cite{Nzar_ACS2016,Abdullah2017} transport.
In this paper, we investigate the properties of resonant current generated due to multi-resonant states 
between the multi-levels QD system and the photons in the cavity. The steady-state resonant current is investigated 
using a Markovian-quantum master equation here. We show that several resonant states can be 
observed enhancing the electron transport. The influence of the photon polarization in a three-dimensional
cavity is demonstrated.

The outline of the paper is as follows. We describe the model system  and transport formalism in~\sec{Sec:Model}.
Results are discussed for the model in~\sec{Sec:Results}. Finally, we draw our conclusion in~\sec{Sec:Conclusion}.

\section{Hamiltonian, transport formalism, and observables}\label{Sec:Model}

\begin{figure}[htb]
\includegraphics[width=0.45\textwidth]{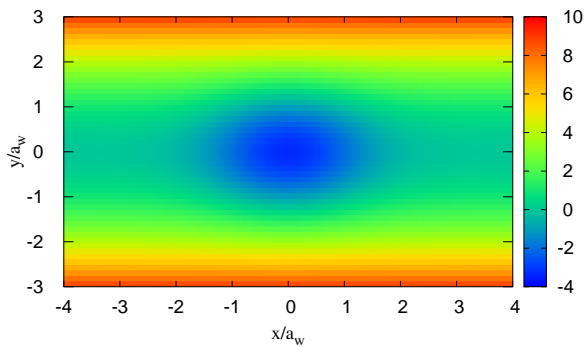}
\caption{The potential $V_r(\mathbf{r})$ defining the central QD system
 that will be coupled diametrically to the semi-infinite left and right leads
 in the $x$-direction.}
\label{fig01}
\end{figure}

We consider a quantum dot with diameter $d \simeq 66.6$~nm embedded in 
a short quantum wire with length $L_x = 150$~nm in the $xy$-plane as is shown in \fig{fig01}.
The QD system is coupled to two electron reservoirs or leads from both ends assuming the 
chemical potential of the left lead $\mu_L$ is higher than that of the right lead $\mu_R$. As a result, 
electrons flow from the left lead to the right lead through the QD system due to the bias voltage 
$e V_{\rm bias} = \mu_L - \mu_R$. The total system, the QD system and the leads, are made of 
GaAs material.

\subsection{Hamiltonian of the QD-Cavity system}

The total Hamiltonian of the QD system coupled to the cavity can be written as
\begin{align}\label{HS}
      H_\mathrm{S} &=  \sum_{n,n^{\prime}} \langle\psi_n \lvert \left[ \frac{(\bm{\pi}_e + 
      \frac{e}{c} \mathbf{A}_{\gamma})^2}{2m^*}
        + V_\mathrm{QD} + eV_\mathrm{g} \right]\lvert\psi_{n^{\prime}}\rangle  d_n^{\dagger} d_{n^{\prime}}
      \nonumber \\
      & + H_{\rm Z} + H_{\mathrm{C}} + H_{\gamma}.
\end{align}
The first line of \eq{HS} displays the QD system without the Coulomb interaction,
where $|\psi_n\rangle$ is a single-electron state (SES), $m^*$ is the electron effective mass,
$e$ is the electron charge and $d^\dagger_{n}$ and $d_{n^{\prime}}$ are the creation and annihilation operators 
of the electron, respectively. 
Furthermore, $\bm{\pi}_e= p+\frac{e}{c}\mathbf{A}_{\mathrm{B}}$, where $p$ is the momentum
operator, $\mathbf{A}_{\mathrm{B}}$ is the magnetic vector potential 
$\mathbf{A}_{\mathrm{B}}$ = ($0,-By,0$) causing an external constant magnetic field,
and $\mathbf{A}_{\gamma}$ is the photon vector potential, that can be defined as 
\begin{equation}
      \mathbf{A_{\gamma}} = A_{\hat{\gamma}}
       \left( a+a^{\dagger} \right)\mathbf{\hat{e}}\, ,
\end{equation}
herein, $A_{\hat{\gamma}}$ is the amplitude of the quantized photons in the cavity,
and $\mathbf{\hat{e}}$ is the unit vector that determines the direction of the photon polarization 
either parallel ($e_x$) in a TE$_{011}$ mode or perpendicular ($e_y$) in a TE$_{101}$ mode
to the transport direction. 
 The electron-photon coupling strength is determined by $g_{\gamma} = e A_{\hat{\gamma}} \Omega_w a_w/c$, 
where $\Omega_w$ the electron effective confinement frequency, and $a_w$ is 
the effective magnetic length~\cite{Vidar85.075306,Nzar_IEEE_2016}.
The QD potential is defined as
\begin{equation}
      V_{\rm QD}(\mathbf{r}) = V_{\rm 0}\; {\rm exp}{(-\gamma_x^2 x^2-\gamma_y^2 y^2)}.
      \label{V_QD}
\end{equation}\\
where $V_{\rm 0}$ determines the depth of the QD, 
and $\gamma_x$ and $\gamma_y$ define the diameter of the QD. $V_{\rm g}$ is the gate voltage 
that moves the energy levels of the QD system with respect to the chemical potential of the leads.

In the second line of \eq{HS}, $H_{\rm Z}= \pm g^{*}\mu_B B/2$ is the Zeeman Hamiltonian 
introducing the interaction between the magnetic moment of an electron and the external magnetic field (B), 
with $\mu_B$ the Bohr magneton and $g^{*} = -0.44$ the 
effective g-factor for GaAs. $H_{\rm C}$ is the electron-electron interaction Hamiltonian
\begin{equation}
 H_{\mathrm{C}} = \frac{1}{2}\sum_{\substack{ nn^{\prime} \\  mm^{\prime}}} 
                  (V_{\mathrm{C}})_{nn',m'm}d_n^\dagger d_{n'}^\dagger d_{m}d_{m'} 
\end{equation}
where $(V_{\mathrm{C}})_{nn',m'm}$ are the Coulomb matrix elements in the SES basis~\cite{PhysRevB.82.195325}, and 
\begin{equation}
 H_{\gamma} = \hbar\omega_{\gamma} a^{\dagger} a 
\end{equation}
refers to the free photon field where $\hbar\omega_{\gamma}$ is the photon energy, 
and $a^{\dagger}(a)$ is the photon creation (annihilation) operator, respectively.
Here we assume the photon wavelength to be much larger than the size of the short
quantum wire with the embedded dot.

\subsection{Transport formalism and observables}

To calculate the evolution of the electrons through the QD system in the steady-state regime, 
we use a non-Markovian master equation with a kernel evaluated up to second order 
in the QD system-lead coupling~\cite{Vidar61.305}. The density operator used in this formalism is uncorrelated
before the time of coupling, $t_0$.
Therefore, the total density operator up to the time of coupling is the tensor product
\begin{equation}
 \hat{\rho}(t_0) = \hat{\rho}_\mathrm{L}(t_0)\hat{\rho}_\mathrm{R}(t_0) \hat{\rho}_{\mathrm{S}}(t_0),
\end{equation}
where $\hat{\rho}_{\mathrm{S}}(t_0)$ is initial value of the density operator of the QD system,
and $\hat{\rho}_{l}(t_0)$ is the density operators of the ($l$) lead with $l$ expressing
the left (L) and the right (R) lead.
Once the QD system and the leads are coupled, the density operators become correlated. 
Here, we consider projection operators that maps the total Liouville operator onto the relevant 
part of the system $P = \rho_l {\rm Tr}_l$. The reduced density operator is then found by 
taking the trace over the Fock space of the electron reservoirs using ${\rm T}_l(\rho_l)= 1$, obtaining
\begin{equation}
 \hat{\rho}_{\mathrm{S}} = \mathrm{Tr}_\mathrm{LR}(\hat{\rho}),
\end{equation}
where the reduced density operator express the state of the electrons in the QD system under 
the effect of the leads. Using the projection formalism of Nakajima and Zwanzig~\cite{Zwanzing.33.1338,Nakajima20.948},
one may get the reduced density operator that demonstrates the transport properties of the QD system 
in the steady-state as~\cite{Vidar61.305,JONSSON201781}   
\begin{align}
      \partial_t\hat{\rho}_\mathrm{S}(t) = &-\frac{i}{\hbar}[\hat{H}_\mathrm{S},\hat{\rho}_\mathrm{S}(t)]
      -\left\{\Lambda^\mathrm{L}[\hat{\rho}_\mathrm{S} ;t]+\Lambda^\mathrm{R}[\hat{\rho}_\mathrm{S} ;t]\right\}\nonumber\\
      &-\frac{\bar{\kappa}}{2\hbar}(\bar{n}_\mathrm{R}+1)\left\{2\alpha\hat{\rho}_\mathrm{S}\alpha^\dagger - \alpha^\dagger \alpha\hat{\rho}_\mathrm{S} 
                                                                                 - \hat{\rho}_\mathrm{S}\alpha^\dagger \alpha\right\}\nonumber\\
       &-\frac{\bar{\kappa}}{2\hbar}(\bar{n}_\mathrm{R})\left\{2\alpha^\dagger\hat{\rho}_\mathrm{S}\alpha  - \alpha\alpha^\dagger\hat{\rho}_\mathrm{S} 
                                                                                 - \hat{\rho}_\mathrm{S}\alpha\alpha^\dagger\right\}.
\label{NZ-eq}
\end{align}
The operators $\Lambda^\mathrm{L}$ and $\Lambda^\mathrm{R}$ describe the ``dissipation'' processes caused by both electron 
reservoirs or leads, 
$\bar{\kappa}$ is the photon decay constant, and $\bar{n}_\mathrm{R}$ is the mean photon number of the photon reservoir.
The second and the third lines in Eq.\ \ref{NZ-eq} embody the photon dissipation of the cavity. 
$\alpha^\dagger$ ($\alpha$) represent the operator in the non-interacting photon number basis, $a^\dagger$ ($a$), transformed 
to the interacting electron photon basis using the rotating wave approximation
~\cite{PhysRevA.31.3761, PhysRev.129.2342, PhysRevA.84.043832, GUDMUNDSSON20181672}.
We have taken care in deriving the damping terms in Eq.\ \ref{NZ-eq} when they are transformed 
from the basis of non-interacting photons to the basis of interacting electrons and photons 
(the eigenstates of $H_\mathrm{S}$ (\ref{HS})).
In the Schr{\"o}dinger picture, this can be done by neglecting all 
high frequency creation terms in the transformed annihilation operators, and all high frequency annihilation 
terms in the transformed creation operators. This guarantees that the open system will evolve into 
the correct physical steady state with respect to the photon decay
\cite{PhysRev.129.2342,PhysRevA.31.3761,PhysRevA.84.043832,PhysRevA.80.053810,PhysRevA.75.013811}.

From the reduced density operator, the physical observables can be calculated. 
For instance, the current from the left lead into the QD system, $I_{\rm L}$,
and the current from it into the right lead, $I_{\rm R}$, can 
be defined as 
\begin{equation}
 I_{l} = {\rm Tr}_\mathrm{S} \Big( \Lambda^{l}[\hat{\rho}_\mathrm{S};t] Q \Big).
\end{equation}
Herein, $Q = -e \sum_i d_i^\dagger d_i$ is the charge operator of the QD system. 
In order to enhance the accuracy of the results we use a step wise numerical
diagonalization and truncation of large many-body Fock spaces \cite{Vidar61.305}.
At the end of that procedure the Markovian master equation is solved in a 
Liouville space constructed from the fully interacting many-body Fock space
including all interactions accounted for in the closed system.

\section{Results}\label{Sec:Results}

In this section, we show the results of our calculation for a QD system coupled to the photon cavity.
The total system, the QD system and the leads, are exposed to a weak external magnetic field $B = 0.1$~T, 
implying the effective magnetic length to be $a_w = 23.8$~nm.
The role of this magnetic field is to lift the spin degeneracy by a small Zeeman splitting,
to avoid numerical difficulties. The bias window is fixed by setting 
$\mu_L = 1.65$~meV, and $\mu_L = 1.55$~meV located between the first and second bands of the leads. 
The gate voltage that moves and positions the energy levels of the QD system with respect the leads
is fixed at $0.651$~meV. In addition the temperature of the leads is constant ($T_l = 0.5$~K). 
The effective confinement energy is $\hbar \Omega_w = \big[ \hbar\Omega_0^2 + \hbar \omega_c^2 \big]^{1/2}$, 
where $\hbar\Omega_0 = 2.0$~meV is the electron confinement energy, and $\hbar \omega_c = 0.172$~meV is the 
cyclotron energy at a given external magnetic field. Furthermore, the electron-photon coupling strength is 
fixed at $g_{\gamma} = 0.1$~meV.

\subsection{$x$-polarized photon field}

First, we consider the photon field being parallel to the direction of electron transport, i.e., the $x$-direction.
\begin{figure}[htb!]
 \includegraphics[width=0.32\textwidth]{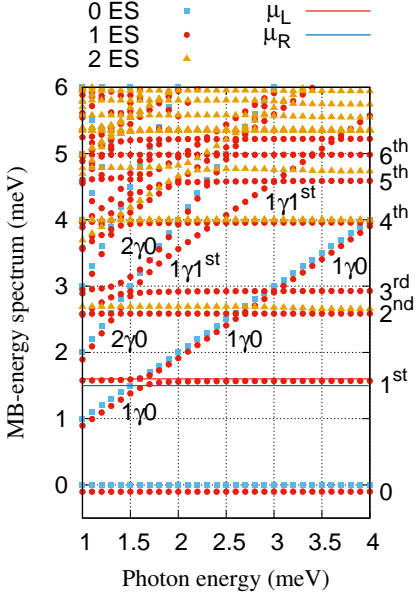}
\caption{Many-Body (MB) energy spectrum ($\rm E_{\rm \mu}$) of the QD system coupled to the cavity versus 
        the photon energy $\hbar \omega_{\gamma}$ for $x$-polarized photon field,
        where 0ES (blue squares) are zero-electron states, 1ES (red circles) are one-electron states, 
        and 2ES (brown triangles) are two-electron states.
        The chemical potential of the left lead is $\mu_L = 1.65$~meV (red line) and the right lead is $\mu_L = 1.55$~meV (blue line).
        0 indicates the one-electron ground-state energy, $1\gamma$0 and $2\gamma$0 refer to the one- and two-photon replica
        of the 0, and 1$^{\rm st}$, 2$^{\rm nd}$, 3$^{\rm rd}$, 4$^{\rm th}$, 5$^{\rm th}$, 6$^{\rm th}$ display the one-electron first-, second-, third-, fourth-, fifth- and sixth-excited state, 
        respectively. The $1\gamma$1$^{\rm st}$ indicates the one-photon replica state of the 1$^{\rm st}$.
        The electron-photon coupling strength $g_{\gamma} = 0.1$~meV, $\bar{\kappa} = 10^{-5}$, and $\bar{n}_\mathrm{R} = 1$.
        The magnetic field is $B = 0.1~{\rm T}$, $V_{\rm g} = 0.651$~meV, $T_{\rm L, R} = 0.5$~K and $\hbar \Omega_0 = 2.0~{\rm meV}$.}
\label{fig02}
\end{figure}
Figure \ref{fig02} presents the many-body (MB) energy spectrum versus the photon energy 
for the QD system coupled to a cavity with an $x$-polarized photon field.
0ES (blue squares) represent zero-electron states, 1ES (red circles) the one-electron states, 
and 2ES (brown triangles) the two-electron states. The labels 0, 1$\gamma$0, and 2$\gamma$0 refer to 
the ground-state, first and second photon replica of the ground-state, respectively, while 
1$^{\rm st}$ and 1$\gamma$1$^{\rm st}$ are the first-excited state and the first photon 
replica of first-excited states, respectively. 
By a photon replica we mean a many-body electron-photon state with a number of photons 
(which is not conserved) close to integers.
Other labels such as 2$^{\rm nd}$, 3$^{\rm rd}$, 4$^{\rm th}$, 5$^{\rm th}$, and 6$^{\rm th}$ 
represent the second-, third-, fourth-, fifth-, and 
sixth-excited states, respectively. We note that each state here includes the Zeeman spin-down 
and the spin-up states that are split due to the small external magnetic field.

Tuning the photon energy several anti-crossings in the energy spectrum are observed, 
such as an anti-crossing between 1$^{\rm st}$ and 1${\gamma}$0 at $\hbar\omega_{\gamma} = 1.7$~meV;
between 2$^{\rm nd}$ and 1${\gamma}$0  at $\hbar\omega_{\gamma} = 2.7$~meV; 
between 3$^{\rm rd}$ and 1${\gamma}$0 at $\hbar\omega_{\gamma} = 3.0$~meV; 
between 2${\gamma}$0 and 3$^{\rm rd}$  at $\hbar\omega_{\gamma} = 1.5$~meV; 
and between 1${\gamma}$1$^{\rm st}$ and 6$^{\rm th}$  at $\hbar\omega_{\gamma} = 3.4$~meV.
Several further anti-crossings in the energy spectrum can be found for the higher lying energy states.

\begin{figure}[htb]
  \includegraphics[width=0.45\textwidth]{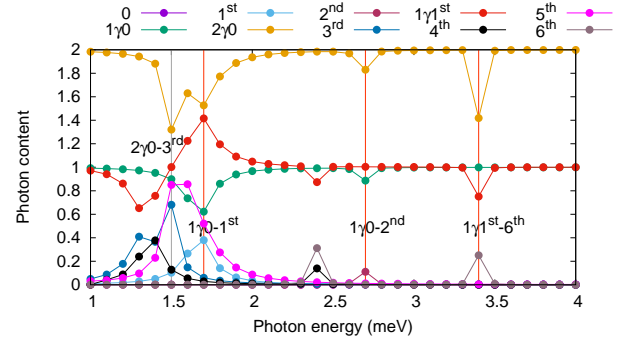}
       \caption{The mean photon number as a function of the photon energy for some lowest states
               of the QD system coupled to a photon field with 
               $x$-polarization. The vertical lines mark
               the location of the main resonance states.
               0 indicates the one-electron ground-state energy (purple), $1\gamma$0 (green) and 
               $2\gamma$0 (brown) refer to the first and second photon replicas of 0, 
               and 1$^{\rm st}$ (light blue), 2$^{\rm nd}$ (dark red), 3$^{\rm rd}$ (dark blue), 4$^{\rm th}$ (black), 5$^{\rm th}$ (magenta), 6$^{\rm th}$ (gray) 
               display the one-electron first-, second-, third-, fourth-, fifth- and sixth-excited state, respectively. 
               The $1\gamma$1$^{\rm st}$ (red) indicates the first photon replica state of the 1$^{\rm st}$.
               The electron-photon coupling strength $\rm g_{\gamma} = 0.1$~meV,  
               $\kappa = 10^{-5}$, and $\bar{n}_\mathrm{R} = 1$.
               The magnetic field is $B = 0.1~{\rm T}$, $V_{\rm g} = 0.651$~meV, 
               and $\hbar \Omega_0 = 2.0~{\rm meV}$.}
\label{fig03}
\end{figure}

Some of the anti-crossings are Rabi-splittings identifiable by the photon-exchange between the 
two anti-crossing states. To find the Rabi-splitting states, \fig{fig03} is presented, where
the photon content or the mean number of photons of some of the lowest and anti-crossing states in the case of 
an $x$-polarized photon field are displayed. 
The vertical lines are the positions in which the main photon-exchange between states occurs.
We start with the photon energy $\hbar \omega_{\gamma} = 1.5$~meV. A photon-exchange of multiple states is observed as follows:
First: the photon-exchange between 3$^{\rm rd}$ (dark blue) and 2$\gamma$0 (brown) (2${\gamma}$0-3$^{\rm rd}$). Second, 
5$^{\rm th}$ (magneta) with 2$\gamma$1$^{\rm st}$  (not shown).
At the photon energy $1.7$~meV, again a photon-exchange of multiple states is seen such as the photon-exchange between 
$1^{\rm st}$ (light blue) and 1$\gamma$0 (green) (1${\gamma}$0-1$^{\rm st}$), and 
between 1$\gamma$1$^{\rm st}$ (red) and 2$\gamma$0 (brown).

Another photon-exchange between 2$^{\rm nd}$ (dark red) and 1$\gamma$0 (green) (1$\gamma$0-2$^{\rm nd}$) at photon energy $2.7$~meV. 
Last, at the photon energy $3.4$~meV the photon-exchange between 1$\gamma$1$^{\rm st}$ (red) and 6$^{\rm th}$ (gray) 
(1$\gamma$1$^{\rm st}$-6$^{\rm th}$) on one hand, and 2$\gamma$0 (brown) with ninth state (now shown) on the other hand 
happen. As the Rabi-splitting between two specific states becomes larger the photon-exchange between 
these two states is enhanced as well. For example, the Rabi-splitting between 1$\gamma$0 and 1$^{\rm st}$ 
at photon energy $1.7$~meV is larger than the splitting between 1$\gamma$0 and 2$^{\rm nd}$ at photon energy $2.7$~meV. 
Therefore the photon-exchange at photon energy $1.7$~meV is larger.  
We should mention that the mean photon number presented here is the photon number of the Zeeman spin-down 
which is exactly the same as spin-up (not shown).

\begin{figure}[h]
  \includegraphics[width=0.45\textwidth,angle=0,bb=70 70 410 250]{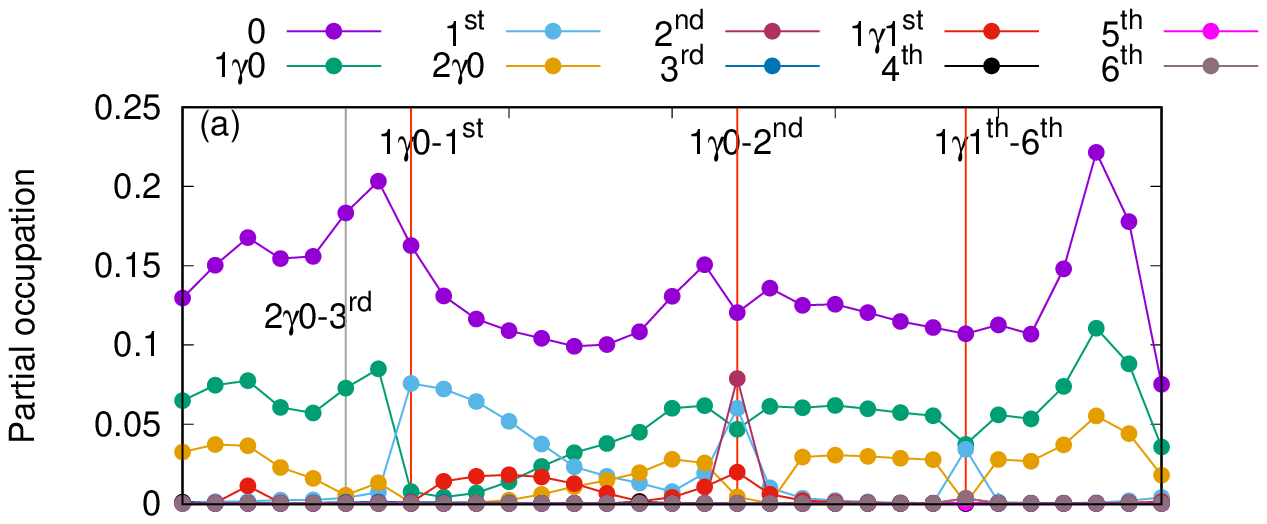}\\
  \includegraphics[width=0.45\textwidth,angle=0,bb=63 55 410 207]{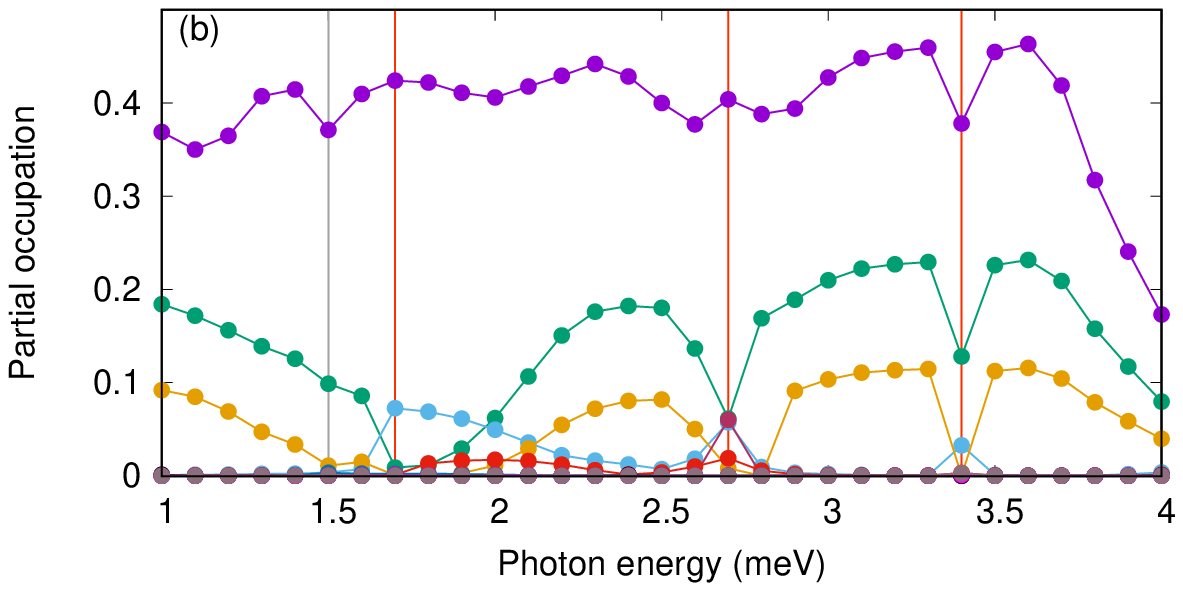}
       \caption{Partial occupation as functions of the photon energy
               for the some lowest states of the QD system with spin down (a) and up (b). 
               The vertical red lines are the location of the main resonance states.
               0 indicates the one-electron ground-state energy (purple), $1\gamma$0 (green) and 
               $2\gamma$0 (brown) refer to the one- and two-photon replica of the 0, 
               and 1$^{\rm st}$ (light blue), 2$^{\rm nd}$ (dark red), 3$^{\rm rd}$ (dark blue), 4$^{\rm th}$ (black), 5$^{\rm th}$ (magenta), 6$^{\rm th}$ (gray) 
               display the one-electron first-, second-, third-, fourth-, fifth- and sixth-excited state, respectively. 
               The $1\gamma$1$^{\rm st}$ (red) indicates the one-photon replica state of the 1$^{\rm st}$.
               The electron-photon coupling strength is $g_{\gamma} = 0.1$~meV, $\bar{n}_\mathrm{R} = 1$, $\bar{\kappa} = 10^{-5}$,
               and the photon field is linearly polarized in the $x$-direction. 
               The chemical potential of the left lead is $\mu_L = 1.65$~meV and the right lead is $\mu_L = 1.55$~meV.
               The magnetic field is $B = 0.1~{\rm T}$, $V_{\rm g} = 0.651$~meV, $T_{\rm L, R} = 0.5$~K, and $\hbar \Omega_0 = 2.0~{\rm meV}$.}
\label{fig04}
\end{figure}
To investigate the transport properties of the selected states demonstrated in \fig{fig03}, 
the partial occupation is presented in \fig{fig04} for Zeeman spin-down (a), and spin-up component (b) in the steady-state.
At the ``low'' photon energy ($1.0$~meV) the 0, 1$\gamma$0, and 2$\gamma$0 are occupied for both 
spin components.

Around the photon energy $1.7$~meV corresponding to the photon-exchange of (1${\gamma}$0-1$^{\rm st}$) (see \fig{fig03}), 
the occupation of 1$\gamma$0 (green) is decreasing to zero and the occupation of 1$^{\rm st}$ (light blue) 
is increased. The reason of charging and discharging of these two states here is caused by the Rabi-splitting effect between them.
Tuning the photon energy to higher value, the 1$\gamma$0 acquires again a charge  while the 1$^{\rm st}$ discharges. 

At the photon energy $2.7$~meV, the occupation of 2$^{\rm nd}$ (dark red) is increased while the occupation of 
1${\gamma}$0 (green) is again decreased which confirms the resonance between these two states.
The same story is also true for the last resonant line at photon energy $3.4$~meV when the occupation of the 6$^{\rm th}$ is 
slightly enhanced.

\begin{figure}[h]
  \includegraphics[width=0.45\textwidth,angle=0,bb=70 70 410 250]{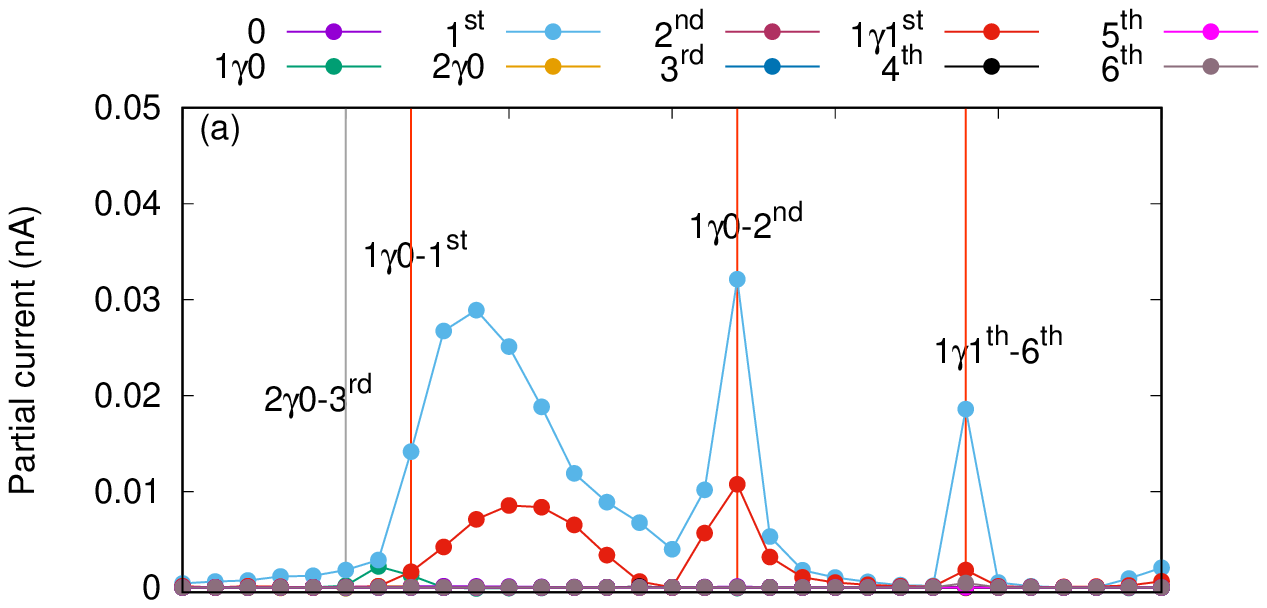}\\
  \includegraphics[width=0.45\textwidth,angle=0,bb=70 55 410 215]{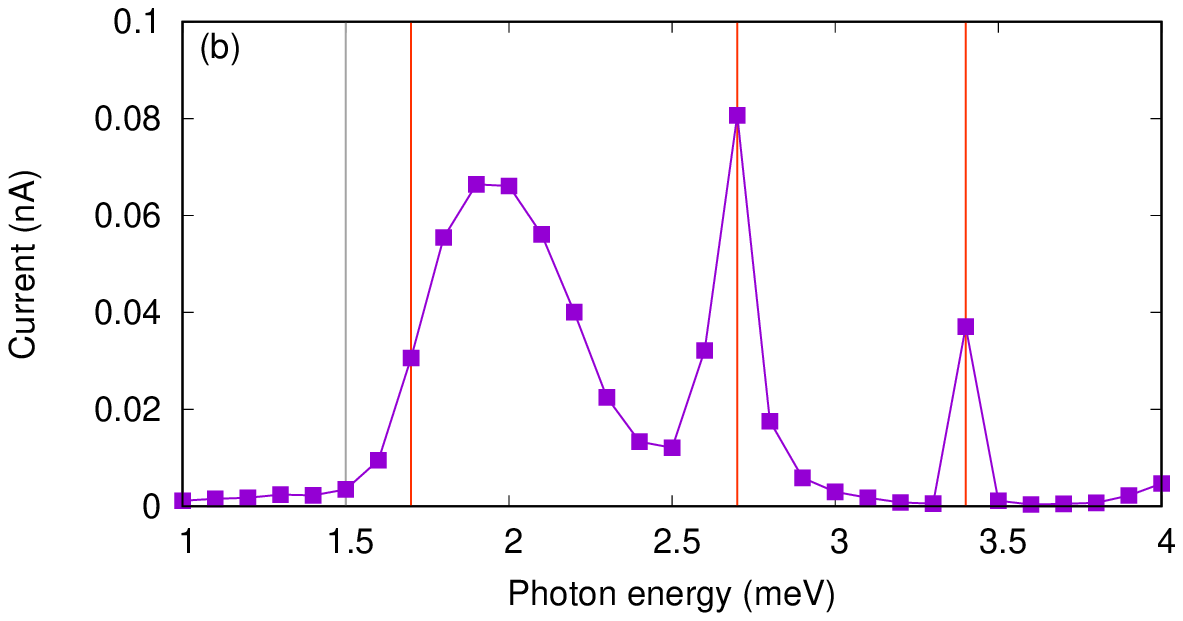}
       \caption{(a) Partial current from the left lead to the QD-system ($I_{\rm L}$)
               as functions of the photon energy for some lowest states with spin down. 
               0 indicates the one-electron ground-state energy (purple), $1\gamma$0 (green) and 
               $2\gamma$0 (brown) refer to the one- and two-photon replica of the 0, 
               and 1$^{\rm st}$ (light blue), 2$^{\rm nd}$ (dark red), 3$^{\rm rd}$ (dark blue), 4$^{\rm th}$ (black), 5$^{\rm th}$ (magenta), 6$^{\rm th}$ (gray) 
               display the one-electron first-, second-, third-, fourth-, fifth- and sixth-excited state, respectively. 
               The $1\gamma$1$^{\rm st}$ (red) indicates the one-photon replica state of the 1$^{\rm st}$.
               (b) Total current including both spin components, down and up, as a function of the photon energy. 
               The electron-photon coupling strength is $g_{\gamma} = 0.1$~meV. $\bar{n}_\mathrm{R} = 1$, $\bar{\kappa} = 10^{-5}$,
               and the photon field is linearly polarized in the $x$-direction. 
               The chemical potential of the left lead is $\mu_L = 1.65$~meV and the right lead is $\mu_L = 1.55$~meV.
               The magnetic field is $B = 0.1~{\rm T}$, $V_{\rm g} = 0.651$~meV, $T_{\rm L, R} = 0.5$~K, and $\hbar \Omega_0 = 2.0~{\rm meV}$.}
\label{fig05}
\end{figure}
For further investigation of the system, the partial current of the lowest states as a function of photon energy is shown in \fig{fig05}(a) 
for the QD system with spin-down. It is observed that the current is maximum around or at the Rabi-resonant states 
of 1${\gamma}$0-1$^{\rm st}$, 1$\gamma$0-2$^{\rm nd}$, and 1$\gamma$1$^{\rm st}$-6$^{\rm th}$. 
The current here is generated due to the bias window, i.e., the location of the bias window tells us which energy states 
should be responsible to the current transport.
The most active states contributing to the current transport are the 1$^{\rm st}$ and 1$\gamma$1$^{\rm st}$ 
because the 1$^{\rm st}$ is located in the bias window. As a result, the photon replicas such as 1$\gamma$1$^{\rm st}$
actively participate in the current transport. We note that the partial current for the spin-up is 
qualitatively similar to the spin-down current, but smaller.
In \fig{fig05}(b) the current, including both spin components, from the left lead to the QD-system ($I_{\rm L}$), 
for the case of an $x$-polarized photon field (purple squares) is shown. 
The currents here confirm that the most active state in the transport is the 1$^{\rm st}$ and it's photon replica states.

We notice that the Rabi-resonant states increase the current in the system, which in turn enhance the efficiency of the 
system.

\subsection{$y$-polarized photon field}

We now assume the photon field is perpendicular to the direction of the electron transport, i.e., $y$-direction.
\begin{figure}[htb!]
  \includegraphics[width=0.32\textwidth]{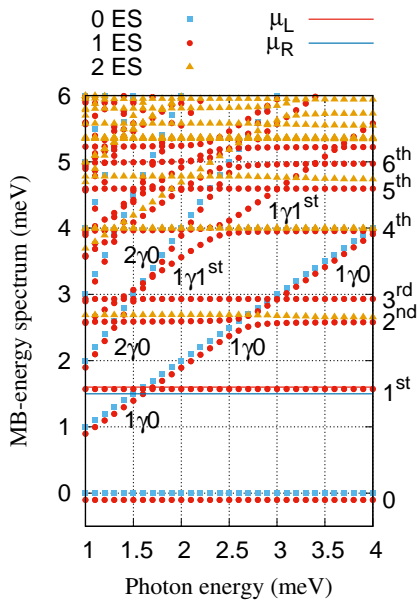}
\caption{Many-Body (MB) energy spectrum ($\rm E_{\rm \mu}$) of the quantum dot system coupled to the cavity versus 
        the photon energy $\hbar \omega_{\gamma}$   for $y$-polarized photon field
        where 0ES (blue squares) are zero-electron states, 1ES (red circles) are one-electron states, 
        and 2ES (brown triangles) are two-electron states.
        The chemical potential of the left lead is $\mu_L = 1.65$~meV (red line) and the right lead is $\mu_L = 1.55$~meV (blue line).
        0 indicates the one-electron ground-state energy, $1\gamma$0 and $2\gamma$0 refer to the one- and two-photon replica
        of the 0, and 1$^{\rm st}$, 2$^{\rm nd}$, 3$^{\rm rd}$, 4$^{\rm th}$, 5$^{\rm th}$, 6$^{\rm th}$ display the 
        one-electron first-, second-, third-, fourth-, fifth- and sixth-excited state, 
        respectively. The $1\gamma$1$^{\rm st}$ indicates the one-photon replica state of the 1$^{\rm st}$.
        The electron-photon coupling strength $\hbar \omega_{\gamma} = 0.1$~meV, $\bar{\kappa} = 10^{-5}$, and $\bar{n}_\mathrm{R} = 1$
        The magnetic field is $B = 0.1~{\rm T}$, $V_{\rm g} = 0.651$~meV, $T_{\rm L, R} = 0.5$~K and $\hbar \Omega_0 = 2.0~{\rm meV}$.}
\label{fig06}
\end{figure}
The energy spectrum of the QD system coupled to the cavity in the $y$-polarized photon is shown in 
\fig{fig06}.

Comparing to the energy spectrum in the case of an $x$-polarization presented in \fig{fig02}, 
the anti-crossing energy gap between 1${\gamma}$0 and 1$^{\rm st}$ at photon energy $1.7$~meV is much smaller, 
while the anti-crossing energy gap between 1${\gamma}$0 and 2$^{\rm nd}$ at photon energy $2.7$~meV is larger here.
This is related to the geometry of charge distribution of the states which is 
more polarized in the $x$-direction~\cite{Nzar.25.465302,PhysicaE.64.254}. 
In addition, the anti-crossing energy gap between 4$^{\rm th}$ and 1${\gamma}$1$^{\rm st}$ at the photon energy $2.4$~meV 
is larger here compared to the $x$-polarized photon field.

\begin{figure}[htb]
  \includegraphics[width=0.45\textwidth]{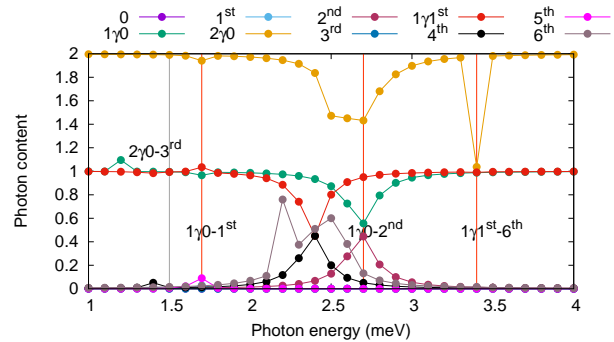}
       \caption{Photon content as a function of the photon energy for some lowest states
               of the QD system coupled to photon field with 
               $y$-polarization. The vertical red lines are 
               the location of the main resonance states.
               0 indicates the one-electron ground-state energy (purple), $1\gamma$0 (green) and 
               $2\gamma$0 (brown) refer to the one- and two-photon replica of the 0, 
               and 1$^{\rm st}$ (light blue), 2$^{\rm nd}$ (dark red), 3$^{\rm rd}$ (dark blue), 4$^{\rm th}$ (black), 5$^{\rm th}$ (magenta), 6$^{\rm th}$ (gray) 
               display the one-electron first-, second-, third-, fourth-, fifth- and sixth-excited state, respectively. 
               The $1\gamma$1$^{\rm st}$ (red) indicates the one-photon replica state of the 1$^{\rm st}$.
               The electron-photon coupling strength $\rm g_{\gamma} = 0.1$~meV,  
               $\kappa = 10^{-5}$, and $\bar{n}_\mathrm{R} = 1$.
               The magnetic field is $B = 0.1~{\rm T}$, $V_{\rm g} = 0.651$~meV, 
               and $\hbar \Omega_0 = 2.0~{\rm meV}$.}
\label{fig07}
\end{figure}

Since the anti-crossing at the photon energy $1.7$~meV is small in the $y$-polarization,
the photon-exchange between the 1${\gamma}$0 (green) and the 1$^{\rm st}$ (light blue) is weak as is shown in \fig{fig07}.
But the anti-crossing at the photon energy $2.7$~meV is large here, the photon-exchange between 
1${\gamma}$0 (green) and the 2$^{\rm nd}$ (dark red) is thus enhanced. 
The photon-exchange between the multiple states in the range of photon energy [$1.0-1.6$]~meV is not seen here, 
while very active photon-exchange between the states in the same range of photon energy were observed
in the $x$-polarized photon field.
In addition, the rate of photon-exchange between 4$^{\rm th}$ (black) and 1${\gamma}$1$^{\rm st}$ (red) at the photon energy $2.4$~meV, 
and 6$^{\rm th}$ (gray) and 2${\gamma}$0 (brown) at the photon energy $2.5$~meV are enhanced which confirm a stronger resonance between 
these states comparing to the $x$-polarization.

\begin{figure}[h]
  \includegraphics[width=0.45\textwidth,angle=0,bb=70 70 410 250]{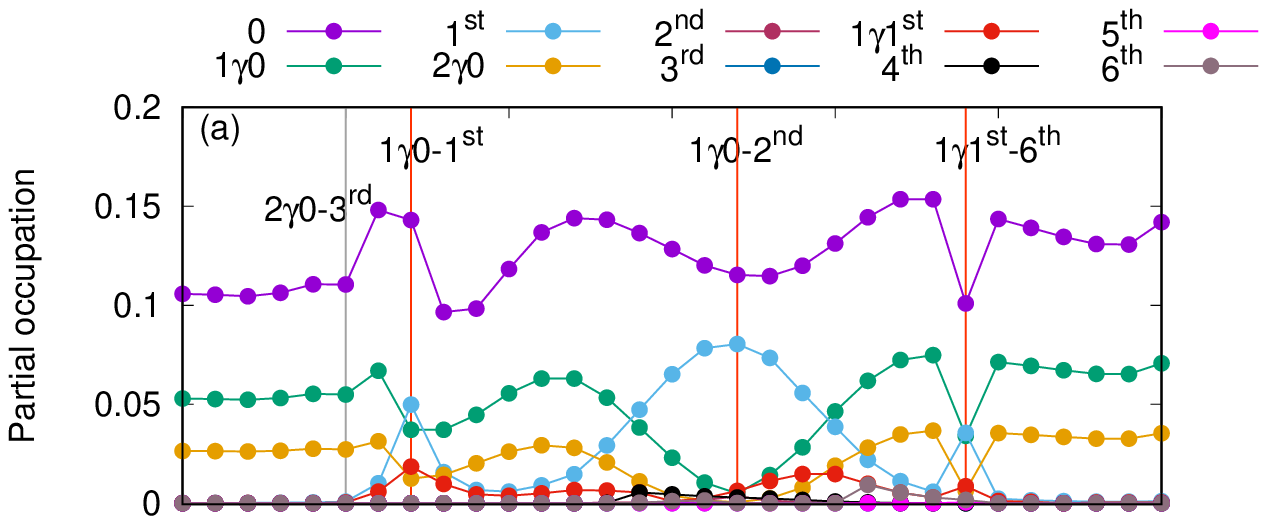}\\
  \includegraphics[width=0.45\textwidth,angle=0,bb=63 55 410 207]{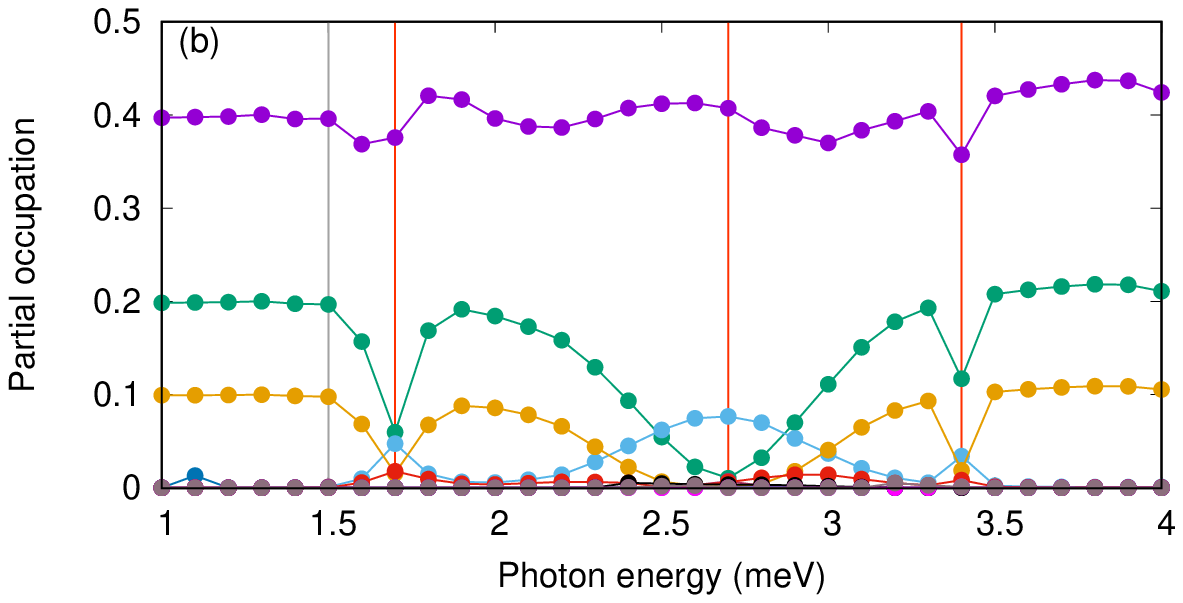}
       \caption{Partial occupation as functions of the photon energy
               for the some lowest states of the QD system coupled to photon field 
               with $y$-polarization in the case of spin-down (a) and spin-up (b). 
               The vertical red lines are the location of the main resonance states.
               0 indicates the one-electron ground-state energy (purple), $1\gamma$0 (green) and 
               $2\gamma$0 (brown) refer to the one- and two-photon replica of the 0, 
               and 1$^{\rm st}$ (light blue), 2$^{\rm nd}$ (dark red), 3$^{\rm rd}$ (dark blue), 4$^{\rm th}$ (black), 5$^{\rm th}$ (magenta), 6$^{\rm th}$ (gray) 
               display the one-electron first-, second-, third-, fourth-, fifth- and sixth-excited state, respectively. 
               The $1\gamma$1$^{\rm st}$ (red) indicates the one-photon replica state of the 1$^{\rm st}$.
               The electron-photon coupling strength is $g_{\gamma} = 0.1$~meV, $\bar{n}_\mathrm{R} = 1$, $\bar{\kappa} = 10^{-5}$,
               and the photon field is linearly polarized in the $y$-direction. 
               The chemical potential of the left lead is $\mu_L = 1.65$~meV and the right lead is $\mu_L = 1.55$~meV.
               The magnetic field is $B = 0.1~{\rm T}$, $V_{\rm g} = 0.651$~meV, $T_{\rm L, R} = 0.5$~K, and $\hbar \Omega_0 = 2.0~{\rm meV}$.}
\label{fig08}
\end{figure}

Figure \ref{fig08} shows the partial occupation of the selected states of 
the QD system with spin-down (a) and spin-up (b) in the case of $y$-polarized of the 
photon field. 
At photon energy $1.7$~meV the 1${\gamma}$0 (green) is discharging and 1$^{\rm st}$ (light blue) is charging. 
The occupation of both states is slightly suppressed compared to what is in the $x$-polarized photon field. 
This is due to the small Rabi-splitting and low photon-exchange between these two states here.
At the other two resonant energies with photon energy $2.7$ and $3.4$~meV, the 1$\gamma$0 is discharging and 
the 1$^{\rm st}$ is occupied indicating that the resonance between 1$\gamma$0 and 2$^{\rm nd}$ at $2.7$~meV and 
the resonance between 1$\gamma$1$^{\rm st}$ and 6$^{\rm th}$ do not play an important 
role in the transport for the $y$-polarized photon field.

\begin{figure}[h]
  \includegraphics[width=0.45\textwidth,angle=0,bb=70 70 410 260]{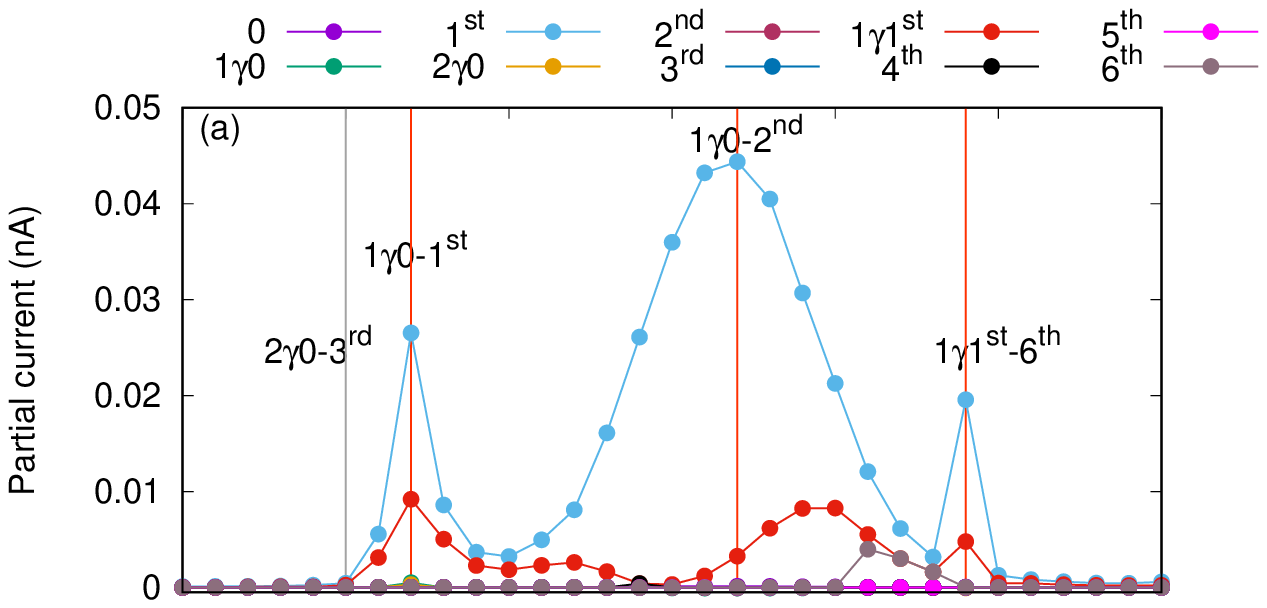}\\
  \includegraphics[width=0.45\textwidth,angle=0,bb=70 55 410 215]{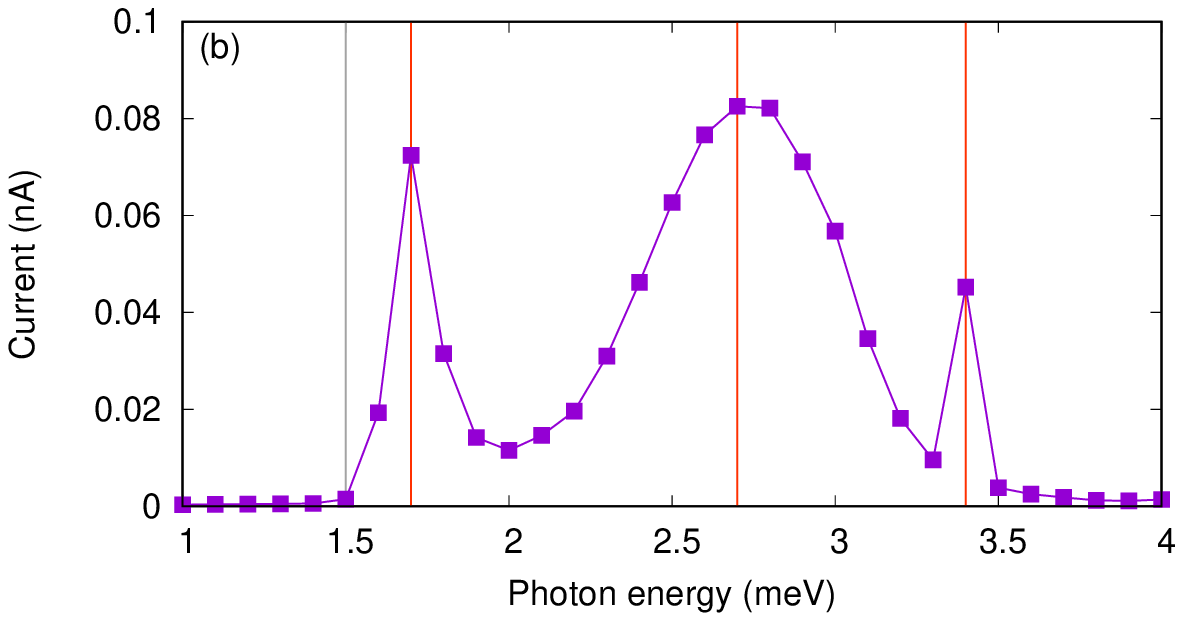}
       \caption{(a) Partial current from the left lead to the QD-system ($I_{\rm L}$)
               as functions of the photon energy for some lowest states with spin down. 
               0 indicates the one-electron ground-state energy (purple), $1\gamma$0 (green) and 
               $2\gamma$0 (brown) refer to the one- and two-photon replica of the 0, 
               and 1$^{\rm st}$ (light blue), 2$^{\rm nd}$ (dark red), 3$^{\rm rd}$ (dark blue), 4$^{\rm th}$ (black), 5$^{\rm th}$ (magenta), 6$^{\rm th}$ (gray) 
               display the one-electron first-, second-, third-, fourth-, fifth- and sixth-excited state, respectively. 
               The $1\gamma$1$^{\rm st}$ (red) indicates the one-photon replica state of the 1$^{\rm st}$.
               (b) Total current including both spin components, down and up, as a function of the photon energy. 
               The electron-photon coupling strength is $g_{\gamma} = 0.1$~meV, $\bar{n}_\mathrm{R} = 1$, $\bar{\kappa} = 10^{-5}$,
               and the photon field is linearly polarized in the $y$-direction. 
               The chemical potential of the left lead is $\mu_L = 1.65$~meV and the right lead is $\mu_L = 1.55$~meV.
               The magnetic field is $B = 0.1~{\rm T}$, $V_{\rm g} = 0.651$~meV, $T_{\rm L, R} = 0.5$~K, and $\hbar \Omega_0 = 2.0~{\rm meV}$.}
\label{fig09}
\end{figure}

The most active state in the current transport in the case of $y$-polarization is again the 1$^{\rm st}$ (light blue)
and it's first photon replica state 1$\gamma$1$^{\rm st}$ (red) as is shown in \fig{fig09}(a) for spin-down component.
The current here is maximum at exactly the ``main'' resonant energies. The characteristic of partial current for the spin-up 
is very similar to the current of the spin-down but it is slightly smaller.

In \fig{fig09}(b) the current, including both spin components, from the left lead to the QD-system ($I_{\rm L}$), 
of the system in the case of $y$-polarization (purple squares) is demonstrated. 
The characteristic of the total current here is qualitatively the same as the current for spin-down and the 
peaks are found at the resonance energy states.
The current here indicates that the most active state in the transport is the 1$^{\rm st}$ and it's photon replica states.
%

\section{Conclusion}\label{Sec:Conclusion}

In this work, we have studied charge and transport current carried by multiple resonant energy states 
through a multi-level QD system coupled to a photon cavity. The charge and current display oscillatory behavior with the
underlying resonances in the many-body energy spectrum. It is shown that the photon polarization 
plays an important role in the transport. When the photon field is polarized in the same direction as the electron motion 
through the QD system ($x$-direction), more resonance states are observed. 
This is caused by the geometry of charge distribution in the QD system that is more polarizable in the $x$-direction,
or in other terms the fact that the quantum wire is parabolically confined in the $y$-direction with a
characteristic energy $\hbar\Omega_0=2.0$ meV, but flat in the $x$-direction except for the lone embedded
quantum dot.
  
In addition, at the resonances the current attains maxima, which in turn 
shows increased efficiency of transport for the QD system when it is in resonance with the cavity photon.
The resonances couple together states with different mean photon number facilitating 
transport through the states available close to the bias window.

\begin{acknowledgments}

This work was financially supported by the Research Fund of the University of Iceland,
the Icelandic Research Fund, grant no.\ 163082-051, 
and the Icelandic Infrastructure Fund. 
The computations were performed on resources provided by the Icelandic 
High Performance Computing Center at the University of Iceland.
NRA acknowledges support from University of Sulaimani and 
Komar University of Science and Technology.
CST acknowledges support from Ministry of Science and
Technology of Taiwan under grant No.\ 106-2112-M-239-001-MY3.

\end{acknowledgments}


%

\end{document}